\documentclass[superscriptaddress,prl,twocolumn]{revtex4}

\usepackage{graphicx}
\begin{document}
\title{Topology Induced Coarsening in Language Games}
\author{Andrea Baronchelli}
\affiliation{Dipartimento di Fisica, Universit\`a ``La Sapienza'' and SMC-INFM,
P.le A. Moro 2, 00185 ROMA, (Italy)}
\author{Luca Dall'Asta}
\affiliation{Laboratoire de Physique Th\'eorique (UMR du CNRS 8627),
B\^atiment 210, Universit\'e de Paris-Sud, 91405 ORSAY Cedex (France)}
\author{Alain Barrat} \affiliation{Laboratoire de Physique Th\'eorique
(UMR du CNRS 8627), B\^atiment 210, Universit\'e de Paris-Sud, 91405
ORSAY Cedex (France)} \author{Vittorio Loreto}
\affiliation{Dipartimento di Fisica, Universit\`a ``La Sapienza'' and SMC-INFM,
P.le A. Moro 2, 00185 ROMA, (Italy)}

\begin{abstract}
We investigate how very large populations are able to reach a global
consensus, out of local "microscopic" interaction rules, in the
framework of a recently introduced class of models of semiotic
dynamics, the so-called Naming Game.  We compare in particular the
convergence mechanism for interacting agents embedded in a
low-dimensional lattice with respect to the mean-field case.  We
highlight that in low-dimensions consensus is reached through a
coarsening process which requires less cognitive effort of the agents,
with respect to the mean-field case, but takes longer to complete. In
$1$-d the dynamics of the boundaries is mapped onto a truncated Markov
process from which we analytically computed the diffusion
coefficient. More generally we show that the convergence process
requires a memory per agent scaling as $N$ and lasts a time
$N^{1+2/d}$ in dimension $d \le 4$ (the upper critical dimension),
while in mean-field both memory and time scale as $N^{3/2}$, for a
population of $N$ agents. We present analytical and numerical
evidences supporting this picture.

\end{abstract}

\maketitle

The past decade has seen an important development of the so-called
Semiotic Dynamics, a new field which studies how conventions (or
semiotic relations) can originate, spread and evolve over time in
populations.  This occurred mainly trough the definition of
Language interaction games~\cite{Steels1997, Kirby} in which a
population of agents is seen as a complex adaptive system which
self-organizes~\cite{MatsenNowak2004} as a result of simple local
interactions (games). The interest of physicists for Language Games
comes from the fact that they can be easily formulated as
non-equilibrium statistical mechanics models of interacting agents: at
each time step, an agent updates its state (among a certain set of
possible states) through an interaction with its neighbors. An
interesting question concerns the possibility of convergence towards a
common state for all agents, which emerges without external global
coordination and from purely local interaction rules
\cite{opinions,Axelrod97,voter}.  In this Letter, we focus on the
so-called Naming Games, introduced to describe the emergence of
conventions and shared lexicons in a population of individuals
interacting with each other by negotiations rules, and study how the
embedding of the agents on a low-dimensional lattice influences the
emergence of consensus, which we show to be reached through a
coarsening process.  The original model~\cite{Steels:1995} was
inspired by a well-known experiment of artificial intelligence called
Talking Heads~\cite{Steels1998}, in which embodied software agents
develop their vocabulary observing objects through digital cameras,
assigning them randomly chosen names and negotiating these names with
other agents.
 
Recently a new minimal version of the Naming Game endowed with
simplified interactions rules~\cite{Baronchelli:2005} has been
introduced, that reproduces the phenomenology of the experiments and
is amenable to analytical treatment.  In this model, $N$ individuals
(or agents) observe the same object, trying to communicate its name
one to the other. The identical agents have at their disposal an
internal inventory, in which they can store an unlimited number of
different names or opinions. At the initial time, all individuals have
empty inventories, with no innate terms. At each time step, the
dynamics consists of a pairwise interaction between randomly chosen
individuals.  Each agent can take part in the interaction as a
``speaker'' or as a ``hearer''. The speaker transmits to the hearer a
possible name for the object at issue; if the speaker does not know a
name for the object (its inventory is empty), it invents a name to be
passed to the hearer\footnote{Each word is associated with a numerical
  label. Thus, the invention of a new word simply corresponds to the
  extraction of a random number.}, while in the case it already knows more
synonyms 
(stored in the inventory), it chooses one of them randomly.  The
hearer's move is deterministic: if it possesses the term pronounced by
the speaker, the interaction is a success, and both speaker and hearer
retain that name as the right one, canceling all the other terms in
their inventories; otherwise, the new name is included in the
inventory of the hearer, without any cancellation.

The mean-field (MF) case has been studied in~\cite{Baronchelli:2005}:
the system initially accumulates a large number of possible names for
the object since different agents (speakers) initially invent
different names and propagate them. Interestingly however, this
profusion of different names leads in the end to an asymptotic
absorbing state in which all the agents share the same name.

Although this model leads to the convergence of all agents to a common
state or ``opinion'', it is interesting to notice the important
differences with other commonly studied models of opinion
formation~\cite{opinions,Axelrod97,voter}.  For example, each agent
can potentially be in an infinite number of possible discrete states
(or words, names), contrarily to the Voter model in which each agent
has only two possible states~\cite{voter}.  Moreover, an agent can
here accumulate in its memory different possible names for the object,
i.e. wait before reaching a decision. Finally, each dynamical step
involves a certain degree of stochasticity, while in the Voter model,
an agent deterministically adopts the opinion of one of its neighbors.

In this Letter, we study the Naming Game model on low-dimensional lattices:
the agents, placed on a regular $d$-dimensional lattice, can interact
only with their $2d$ nearest neighbors. Numerical and analytical
investigations allow us to highlight important differences with the
mean-field case, in particular in the time needed to reach consensus,
and in the effective size of the inventories or total memory
required. We show how the dynamics corresponds to a coarsening of
clusters of agents sharing a common name; the interfaces between such
clusters are composed by agents who still have more than one possible
name.

\begin{figure} 
\centerline{
\includegraphics*[width=0.35\textwidth]{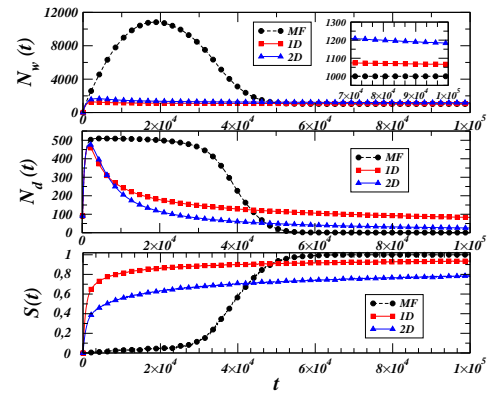}
}
\caption{Time evolution in mean-field and finite dimensions of the
total number of words (or total used memory), for the number of
different words in the system, and for the average success
rate. $N=1024$, average over $1000$ realizations.  The inset in the
top graph shows the very slow convergence in finite dimensions.  }
\label{ng1d2d}
\end{figure}

Relevant quantities in the study of Naming Games are the total number of
words in the system $N_w(t)$, which corresponds to the total memory
used by the agents, the total number of different words $N_d(t)$, and
the average rate of success $S(t)$ of the
interactions. Fig.~\ref{ng1d2d} displays the evolution in time of
these three quantities for the low-dimensional models, compared to the
mean-field case.

In the initial state, all inventories are empty. At short times
therefore, each speaker with an empty inventory has to invent a name
for the object, and many different words are indeed invented. In this
initial phase, the success rate is equal to the probability that two
agents that have already played are chosen again: this rate is proportional to 
$t/E$ where $E$ is the number of possible interacting pairs,
i.e. $N(N-1)/2$ for the mean-field case and $Nd$ in finite
dimensions. $S(t)$ grows thus $N$ times faster in finite dimensions,
as confirmed by numerics. At larger times however, the eventual
convergence is much slower in finite dimensions.

\begin{figure} 
\centerline{
\includegraphics*[width=0.35\textwidth]{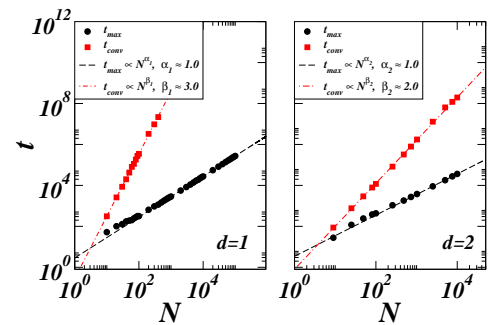} }
\caption{
Scaling of the time at which the number of words is maximal, and
of the time needed to obtain convergence, in $1$ and $2$ dimensions.
}
\label{scaling}
\end{figure}

The curves for $N_w(t)$ and $N_d(t)$ display in all cases a sharp
increase at short times, a maximum for a given time $t_{max}$ and then
a decay towards the consensus state in which all the agents share the
same unique word, reached at $t_{conv}$.  The short time regime
corresponds to the creation of many different words by the
agents. After a time of order $N$, each agent has played typically
once, and therefore ${\cal O}(N)$ different words have been invented
(in fact, typically $N/2$): the total number of {\em distinct} words
in the system grows and reaches a maximum scaling as $N$. Due to the
interactions, the agents accumulate in memory the words they have
invented and the words that other agents have invented. In MF, each
agent can interact with all the others, so that it can learn many
different words, and in fact the maximal memory necessary for each
agent scales as $N^\alpha_{MF}$ with
$\alpha_{MF}=0.5$~\cite{Baronchelli:2005}, so that the total memory
used at the peak is $\sim N^{1.5}$, with many words shared by many
agents, and $t_{max}\sim N^\beta_{MF}$ with $\beta_{MF}=1.5$.
Moreover, during this learning phase, words are not eliminated ($S(t)$
is very small) so that the total number of distinct words displays a
long plateau. The redundancy of words then reaches a sufficient level
to begin producing successful interactions and the decrease of the
number of words is then very fast, with a rapid convergence to the
consensus state. In contrast in finite dimensions words can only
spread locally, and each agent has access only to a finite number of
different words. The total memory used scales as $N$, and the time
$t_{max}$ to reach the maximum number of words in the system scales as
$N^{\alpha_d}$ with $\alpha_1=\alpha_2=1$ (Fig.~\ref{scaling}). No
plateau is observed in the total number of distinct words since
coarsening of clusters of agents soon start to eliminate words.

Furthermore, the time needed to reach consensus, $t_{conv}$, grows as
$N^{\beta_{d}}$ with $\beta_{1} \simeq 3$ in $d=1$ and $\beta_{2}
\simeq 2$ in $d=2$, while $\beta_{MF} \simeq 1.5$
(Fig.~\ref{scaling}). We will now see how such behaviors emerge from a
more detailed numerical and analytical analysis of the dynamical
evolution.

%

\begin{figure} 
\centerline{
\includegraphics*[width=0.35\textwidth]{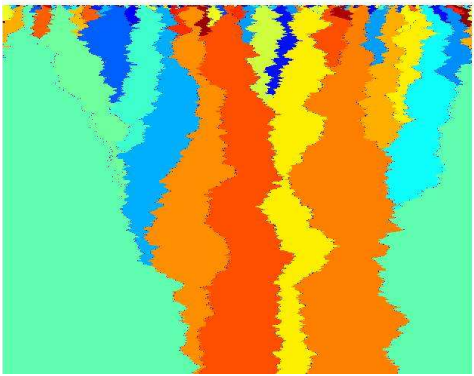}
}
\caption{ Typical evolution of a one-dimensional system
($N=1000$). Black color corresponds to interfaces (sites with more
than one word). The other colors identify different single state
clusters. The vertical axis represents the time ($1000 \times N$
sequential steps), the one-dimensional snapshots are reported on the
horizontal axis.  }
\label{conf1d}
\end{figure}

Fig.\ref{conf1d} reports a typical evolution of agents on a
one-dimensional lattice, by displaying one below the other a certain
number of (linear) configurations corresponding to successive equally
separated temporal steps. Each agent having one single word in memory
is presented by a colored point while agents having more than one word
in memory are shown in black.  This figure clearly shows the growth of
clusters of agents having one single word by diffusion of interfaces
made of agents having more than one word in memory. The fact that the
interfaces remain thin is however not obvious a priori: an agent
having e.g. two words in memory can propagate them to its neighbors,
leading to possible clusters of agents having more than one word.

%


In order to rationalize and quantify such evolution, we consider a
single interface between two linear clusters of agents: in each
cluster, all the agents share the same unique word, say $A$ in the
left-hand cluster and $B$ in the other. The interface is a string of
length $m$ composed of sites in which both states $A$ and $B$ are
present. We call $C_{m}$ this state ${(A+B)}^m$. A $C_{0}$ corresponds
to two directly neighboring clusters ($\cdots AAABBB\cdots$), while
$C_{m}$ means that the interface is composed by $m$ sites in the state
$C=A+B$ ($\cdots AAAC\cdots CBBB\cdots$).  Note that, in the actual
dynamics, two clusters of states $A$ and $B$ can be separated by a
more complex interface. For instance a $C_{m}$ interface can break
down into two or more smaller sets of $C$-states spaced out by $A$ or
$B$ clusters, causing the number of interfaces to grow.  Numerical
investigation shows that such configurations are however eliminated in
the early times of the dynamics.

\begin{figure} 
\centerline{
\includegraphics*[width=0.3\textwidth,angle=-90]{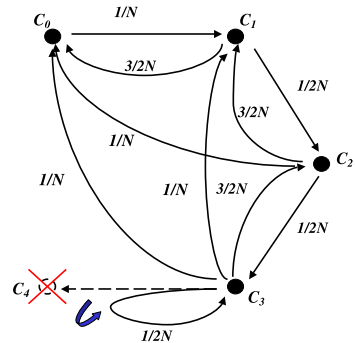} }
\caption{Truncated Markov process associated with interface width dynamics -
  schematic evolution of a $C_0$ interface $\cdots AAABBB\cdots$, cut at
  the maximal width $m=3$.}
\label{markov}
\end{figure}
%

Bearing in mind these hypotheses, an approximate expression for the
stationary probability that two neighboring clusters are separated by
a $C_{m}$ interface can be computed in the following way. In a
one-dimensional line composed of $N$ sites and initially divided into
two clusters of $A$ and $B$, the probability to select the unique
$C_{0}$ interface is $1/N$, and the interacting rules say that the
only possible product is a $C_{1}$ interface. Thus, there is a
probability $p_{0,1}=1/N$ that a $C_{0}$ interface becomes a $C_{1}$
interface in a single time step, otherwise it stays in $C_{0}$. From
$C_{1}$ the interface can evolve into a $C_{0}$ or a $C_{2}$ interface
with probabilities $p_{1,0} = \frac{3}{2 N}$ and $p_{1,2} = \frac{1}{2
N}$ respectively.  This procedure is easily extended to higher values
of $m$. The numerics suggest that we can safely truncate this study at
$m \leq 3$.  In this approximation, the problem corresponds to
determine the stationary probabilities of the Markov chain reported in
Fig.\ref{markov} and defined by transition matrix
\begin{equation}
\mathcal{M} =\left(
\begin{array}{cccc} 
\frac{N-1}{N} & \frac{1}{N} & 0 & 0 \\
\frac{3}{2 N} & \frac{N-2}{N} & \frac{1}{2 N} & 0 \\
\frac{1}{N} & \frac{3}{2 N} & \frac{N-3}{N} & \frac{1}{2 N} \\
\frac{1}{N}  & \frac{1}{N} & \frac{3}{2 N} & \frac{N-4}{N}+\frac{1}{2 N} \\  
\end{array} \right),  
\end{equation}
in which the basis is $\{C_{0}, C_{1}, C_{2}, C_{3}\}$ and the
contribution $\frac{1}{2 N}$ from $C_{3}$ to $C_{4}$ has been
neglected (see Fig.\ref{markov}).  The stationary probability vector
${\bf{P}} = \{P_{0}, P_{1}, P_{2}, P_{3}\}$ is computed by imposing
${\bf{P}} (t+1) - {\bf{P}} (t) = 0$, i.e. $(\mathcal{M}^{T} - I)
{\bf{P}} = 0$, that gives $P_{0} = 133/227 \approx 0.586$, 
$P_{1}= 78/227 \approx 0.344$, $P_{2} = 14/227\approx 0.062$, 
$P_{3} = 2/227 \approx 0.0088$. Direct numerical simulations of the
evolution of a line $\cdots AAABBB\cdots$ yields 
$P_{0} \simeq 0.581$, 
$P_1=0.344$,
$P_2=0.063$,
$P_3=0.01$,
thus clearly confirming the correctness of our approximation.

Since our analysis shows that the width of the interfaces remains
small, we assume that they are punctual objects localized around their
central position $x$: in the previously analyzed case, denoting by
$x_{l}$ the position of the right-most site of cluster $A$ and by
$x_{r}$ the position of the left-most site of cluster $B$, it is given
by $x =\frac{x_{l}+x_{r}}{2}$. An interaction involving sites of an
interface, i.e. an interface transition $C_{m} \rightarrow C_{m'}$,
corresponds to a set of possible movements for the central position
$x$. The set of transition rates are obtained by enumeration of all
possible cases: denoting by $W(x \rightarrow x\pm\delta)$ the
transition probability that an interface centered in $x$ moves to the
position $x\pm\delta$, in our approximation only three symmetric
contributions are present. We obtain $W(x \rightarrow x\pm\frac{1}{2})
= \frac{1}{2N} P_{0} + \frac{1}{N} P_{1} + \frac{1}{N} P_{2} +
\frac{1}{2N} P_{3}$, $W(x \rightarrow x\pm 1) = \frac{1}{2N} P_{2} +
\frac{1}{2N} P_{3}$, $W(x \rightarrow x\pm\frac{3}{2}) = \frac{1}{2N}
P_{3}$.  Using the expressions for the stationary probability $P_{0},
\dots, P_{3}$, we finally get $W(x \rightarrow x\pm\frac{1}{2}) =
\frac{319}{454N}$, $W(x \rightarrow x\pm 1) = \frac{8}{227N}$, and
$W(x \rightarrow x\pm\frac{3}{2}) = \frac{1}{227N}$.

The knowledge of these transition probabilities allows us to write the
master equation for the probability $\mathcal{P}(x,t)$ to find the
interface in position $x$ at time $t$, which, in the limit of
continuous time and space (i.e. writing $\mathcal{P}(x,t+1) -
\mathcal{P}(x,t) \approx \delta t \frac{\partial \mathcal{P}}{\partial
t}(x,t)$, while $\mathcal{P}(x+ \delta x, t) \approx \mathcal{P}(x,t)
+ \delta x \frac{\partial \mathcal{P}}{\partial x}(x,t) +
\frac{{(\delta x)}^2}{2} \frac{{\partial}^2 \mathcal{P}}{{\partial
x}^2}(x,t)$), reads $\frac{\partial \mathcal{P} (x,t)}{\partial t} =
\frac{D}{N} \frac{{\partial}^2 \mathcal{P} (x,t)}{{\partial x}^2}$,
where $D = 401/1816 \simeq 0.221$ is the diffusion coefficient (in the
appropriate dimensional units ${(\delta x)}^2 / \delta t$).

%
\begin{figure} 
\centerline{ \includegraphics*[width=0.33\textwidth]{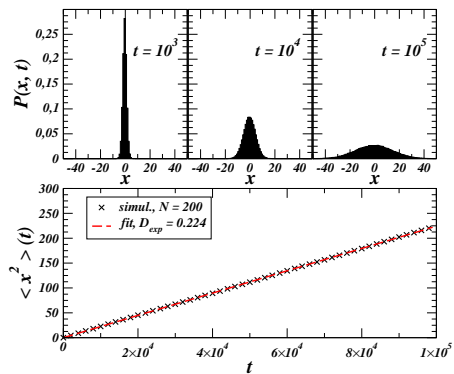} }
\caption{Evolution of the position of an interface $\cdots
AAABBB\cdots$.  Top: evolution of the distribution $\mathcal{P}
(x,t)$. Bottom: evolution of the mean square displacement, showing a
clear diffusive behavior $\langle x^2\rangle = 2 D_{exp} t/N$ with a
coefficient $D_{exp}\approx 0.224$ in agreement with the theoretical
prediction.  }
\label{rw1D}
\end{figure}

These results are confirmed by numerical simulations as illustrated in
Fig.~\ref{rw1D} where the numerical probability $\mathcal{P}(x,t)$ is
shown to be a Gaussian around the initial position, while the
mean-square distance reached by the interface at the time $t$ follows
the diffusion law $\langle x^2 \rangle = 2 D_{exp} t/N$ with $D_{exp}
\simeq 0.224 \approx D$. The dynamical evolution of the Naming Game on
a one-dimensional lattice can then be described as follows: at short
times, pairwise interactions create ${\cal O}(N)$ small clusters,
divided by thin interfaces (see the first lines in
Fig.\ref{conf1d}). We can estimate the number of interfaces at this
time with the number of different words in the lattice, that is about
$N/2$.  The interfaces then start diffusing. When two interfaces meet,
the cluster situated in between the interfaces disappears, and the two
interfaces coalesce.  Such a coarsening leads to the well-known growth
of the typical size $\xi$ of the clusters as $t^{1/2}$. The density of
interfaces, at which unsuccessful interactions can take place, decays
as $1/\sqrt{t}$, so that $1-S(t)$ also decays as
$1/\sqrt{t}$. Moreover, starting from a lattice in which all agents
have no words, a time $N$ is needed to reach a size of order $1$, so
that in fact $\xi$ grows as $\sqrt{t/N}$ (as also shown by the fact
that the diffusion coefficient is $D/N$), which explains the time
$t_{conv} \sim N^3$ needed to reach consensus, i.e. $\xi=N$.

This framework can be extended to the case of higher dimensions. The
interfaces, although quite rough, are well defined and their width
does not grow in time, which points to the existence of an effective
surface tension.  The numerical computation of equal-time pair
correlation function in dimension $d=2$ (not shown) indicates that the
characteristic length scale $\xi$ grows as $\sqrt{t/N}$ (a time ${\cal
O}(N)$ is needed to initialize the agents to at least one word and
therefore to reach a cluster size of order $1$), in agreement with
coarsening dynamics for non-conserved fields~\cite{bray}.  Since
$t_{conv}$ corresponds to the time needed to reach $\xi = N^{1/d}$, we
can argue $t_{conv} \sim N^{1+2/d}$, that has been verified by
numerical simulations in $d=2$ and $d=3$. This scaling and the
observed coarsening behavior suggest that the upper critical dimension
for this system is $d=4$~\cite{bray}.

In conclusion, the study of the low-dimensional Naming Game using
statistical physics methods provides a deeper understanding of the
macroscopical collective dynamics of the model. We have shown how it
presents a very different behavior in low-dimensional lattices than in
mean-field, indicating the existence of a finite upper-critical
dimension.  Low-dimensional dynamics is initially more effective, less
memory per node is required, preventing agents from learning a large
part of the many different words created.  The dynamics then proceeds
by the growth of clusters by coarsening, yielding a slow convergence
to consensus.  In contrast with other models of opinion dynamics
(e.g. the Voter model~\cite{voter2,dornic}), the Naming Game presents
an effective surface tension that is reminiscent of the
non-equilibrium zero-temperature Ising model~\cite{bray}.  In this
respect, it seems interesting to investigate the dynamics of the
Naming Game in other topologies, such as complex networks in which
each node have a finite number of neighbors combined with "long-range"
links \cite{inprep}.  {\bf Acknowledgments}: The authors thank
E. Caglioti, M. Felici and L. Steels for many enlightening
discussions.  A. Baronchelli and V. L. are partially supported by the
EU under contract IST-1940 (ECAgents).  A. Barrat and L.D. are
partially supported by the EU under contract 001907 (DELIS).


\begin{thebibliography}{100}



\bibitem{Steels1997} L. Steels,
Evolution of Communication {\bf 1}, 1-34 (1997).

\bibitem{Kirby} S. Kirby, 
Artificial Life {\bf 8}, 185-215 (2002).

\bibitem{MatsenNowak2004} F. Matsen and M.A. Nowak,
Proc. Natl. Acad. Sci. USA {\bf 101}, 18053-18057 (2004). 

\bibitem{opinions} K. Sznajd-Weron and J. Sznajd,
Int. J. Mod. Phys. C, {\bf 11}, 1157 (2000); G. Deffuant, D. Neau,
F. Amblard and G. Weisbuch, Adv. Compl. Syst. {\bf 3}, 87 (2000);
R. Hegselmann and U. Krause, J. Art. Soc. Soc. Sim. {\bf 5}, issue 3,
paper 2 (2002); P. L. Krapivsky and S. Redner, Phys. Rev. Lett. {\bf
90}, 238701 (2003).

\bibitem{Axelrod97} R. Axelrod, J. of Conflict Resolut., {\bf 41},
203, (1997).

\bibitem{voter}
P.L. Krapivsky,
Phys. Rev. A {\bf 45}, 1067 (1992).

\bibitem{Steels:1995} L. Steels, Artificial Life Journal {\bf 2}, 319
(1995).

\bibitem{Steels1998} L. Steels, 
Autonomous Agents and Multi-Agent Systems {\bf 1}, 169-194 (1998).

\bibitem{Baronchelli:2005}
A. Baronchelli, M. Felici, E. Caglioti, V. Loreto and L. Steels,
arxiv:physics/0509075 (2005).

\bibitem{bray} A. Bray, Adv. in Phys {\bf 51}, 481 (2002).

\bibitem{voter2} E. Ben-Naim, L. Frachebourg, and P. L. Krapivsky,
Phys. Rev. E {\bf 53}, 3078-3087 (1996).

\bibitem{dornic} I. Dornic, H. Chat\'e, J. Chave, and H. Hinrichsen,
Phys. Rev. Lett. {\bf 87}, 045701 (2001).

\bibitem{inprep} L. Dall'Asta et al., work in preparation.

\end{thebibliography}
\end{document}